\documentclass{osa-article}
\journal{osac}
\articletype{Research Article}

\usepackage{siunitx}
\usepackage{bm}
\usepackage{float}
\usepackage{afterpage}
\usepackage{comment}
\usepackage{amsmath}

\begin{document}

\title{Optical transmission matrix measurement sampled on a dense hexagonal lattice}

\author{Pritam Pai,\authormark{*} Jeroen Bosch,\authormark{} and Allard P. Mosk\authormark{}}

\address{\authormark{}Debye Institute for Nanomaterials Science, Utrecht University, P.O. Box 80000, 3508 TA Utrecht, The Netherlands}

\email{\authormark{*}p.pai@uu.nl} 

\begin{abstract}
The optical transmission matrix (TM) characterizes the transmission properties of a sample. We show a novel experimental procedure for measuring the TM of light waves in a slab geometry based on sampling the light field on a hexagonal lattice at the Rayleigh criterion. Our method enables the efficient measurement of a large fraction of the complete TM without oversampling while minimizing sampling crosstalk and the associated distortion of the statistics of the matrix elements. The procedure and analysis described here is demonstrated on a clear sample which serves as an important reference for other systems and geometries such as dense scattering media.
\end{abstract}

\section{Introduction}
In wave physics, the scattering operator of a sample links the incident waves to the ballistic and scattered outgoing waves~\cite{vesperinas_wolf86, Lagendijk1996}. By definition it considers only waves that propagate to, or from, the far field. If the sample has a slab geometry, the part of the scattering operator that relates the incident field to the transmitted one is called the transmission operator (TO). By choosing a basis of incident and outgoing modes one can construct a matrix representation of the transmission operator known as the transmission matrix (TM)~\cite{beenakker, popoff2010, allard_review, rossum, rotter_gigan_review, vPutten_mosk}. A TM measurement that gives full information on the transmission operator requires a basis that spans all incident and outgoing modes of the wave field. The number of independent modes $N_s$ incident on an area $A$ on the surface of a slab is equal to the number of modes in a waveguide of the same area, $N_s = 2\pi A / \lambda^2$~\cite{allard_review, saleh_teich}, where $\lambda$ is the wavelength. It is however not straightforward to find an orthogonal and complete basis on a finite sample in free space.

One of the reasons to measure the TM is to explore the distribution and properties of its singular values which, if the TM accurately represents the TO, correspond to the transmission eigenchannels of the sample. These channels are useful in imaging and communication~\cite{allard_review, Miller19} and are an important building block of quantum transport theory in scattering samples~\cite{nazarov_blanter_2009, economouPRL, Baranger1991a,fisher_lee1981, pendry1990}. The predicted theoretical distribution of the singular values in a multiple scattering waveguide is a bimodal distribution known as the Dorokhov-Mello-Pereyra-Kumar curve~\cite{DMPK, martin_landauer} and the properties of transmission channels are a subject of recent intense research~\cite{vellekoop2008prl, shi2012, Davy13, aubry2014, Hao2014, Liew2014,kim2015,Davy2015,Daniel2016, akbulutPRA, hsu2017, Hong2018, rotter_gigan_review, Fang2019_prb, yilmaz_prl19, yilmaz19, Miller19}. In an open optical system, the finite field of view and finite numerical aperture of the optics and the escape of energy parallel to the surface of the sample make it impossible to measure the complete matrix. The statistics of a partial transmission matrix has been shown to differ strongly from that of a full one~\cite{stone2013, yu2013}. This is less of a problem for guided waves such as microwaves~\cite{shi2012, Davy13}, ultrasound~\cite{sprik2008, aubry2009} and elastic waves~\cite{aubry2014} where it is easier to access a large fraction of the scattered field. It is an ongoing challenge to accurately measure the largest possible fraction of the TM of a three-dimensional sample in an open optical system~\cite{vPutten_mosk}.

Popoff and coauthors published the first measurements of the optical TM of a multiple scattering sample, using a spatial light modulator (SLM) to project light fields on a Hadamard basis and with an unmodulated part of the light as the phase reference~\cite{popoff2010}. They demonstrated transmitting predetermined focused spots and images~\cite{popoff_imaging2010}. The low sampling density in their experiment precluded the observation of open channels or other mesoscopic correlations. Other methods employed a basis of spots in real or Fourier space, using a scan-mirror to enhance measurement speed as well as a separate reference arm to retrieve the transmitted field with a single shot method~\cite{choi2011, kim2012, yu2013, tom_phase_microscopy, akbulutPRA}. Interferometric methods are sensitive to phase drifts, and to this end it has been shown that the TM can also be measured without a reference beam~\cite{Dremeau15}. However, this method requires a much larger number of measurements than the number of independent modes. Optical TM measurements close to the critical sampling density have been performed by scanning the beam on a square lattice~\cite{akbulutPRA, yu2013, cizmar, yilmaz19, yilmaz_prl19}.
When approaching the critical sampling density, where the number of sampled incident modes equals the total number of linearly independent modes $N_s$, bandwidth-limited input spots become correlated as the nearest neighbor spot wavefunctions (usually Airy disks) show increasing overlap, leading to distorted statistics of the TM. The amount of overlap depends on the lattice configuration.

In this paper, we demonstrate TM measurements using a hexagonal lattice of Airy disks spaced at the Rayleigh criterion, where the field overlap between neighboring spots is exactly zero and only small overlap exists between more distant spots. We note that measurements on an undersampled hexagonal grid far from the Rayleigh criterion have been demonstrated in Ref.~\cite{Boniface19}. We show that the hexagonal grid leads to a better representation of the statistics of the TM than square-grid sampling only close to the Rayleigh criterion.
As in~\cite{yu2013,akbulutPRA}, we measure a polarization-complete TM. The spots are scanned in an outward hexagonal spiral which helps determine efficiently the point after which the beam escapes the microscope objective's field of view or numerical aperture. We describe the experimental method and data analysis and perform a measurement of the TM of a  zero-thickness medium, i.e., a clear medium where the focal planes of the input and output optics are the same. In this case the incident and transmitted fields are identical and the transmission operator is diagonal. Crucially, the orthogonality of the basis determines whether the transmission matrix is also diagonal, and we can check directly how the statistics of the TM are influenced by the sampling lattice.

\section{Sampling the transmission matrix}
\label{theory}

Theoretically, the scattering matrix is expressed using a basis set of orthogonal flux-normalized far-field modes. We have natural basis sets in the case of a sample in a waveguide (flux-normalized waveguide modes) as well as in the case of a small scatterer in free space (flux-normalized partial waves)~\cite{Miller19}. For a slab-type sample in free space the S-matrix is conventionally written as $S_{\sigma \bm{k} \sigma' \bm{k'}}$, where $\bm{k}$ and $\bm{k'}$ are the wavevectors and $\sigma$ and $\sigma'$ the polarizations corresponding to the incoming and scattered waves respectively~\cite{vesperinas_book, vesperinas_wolf86, Wolf_1959}. The scattering matrix element $S_{\sigma \bm{k} \sigma' \bm{k'}}$ represents the amplitude scattered into the flux-normalized channel $\bm{k'}$ with polarization $\sigma'$ due to an incident flux with normalized amplitude in channel $\bm{k}$ and polarization $\sigma$. If $k$ and $k'$ are on opposite sides of a slab-type sample we speak of a transmission matrix element. On a finite sample, infinite plane waves are not a useful basis set and one has to resort to modes that are confined both in real space and Fourier space, such as diffraction limited spots.

Here we choose a real-space basis of diffraction limited spots that are generated by our microscope objective. We note that the TM can equivalently be measured in a $k$-space basis, where similar considerations apply.
An objective fulfilling the Abbe sine condition~\cite{abbe_sine} filled with a uniform collimated beam produces an Airy disk in the focal plane, which is a band-limited field defined by the circular 
shape of the pupil. The optimal sampling configuration for band-limited signals by a periodic lattice on a 2D plane is a hexagonal lattice (also known as a triangular lattice)~\cite{hex_optimal_sampling}. If the density of sampling points is at least the critical or Nyquist density, corresponding to one sampling point per mode, the entire signal can be reconstructed from the sampled points. 

\begin{figure}[tb]
	\centering\includegraphics[width=0.67\textwidth]{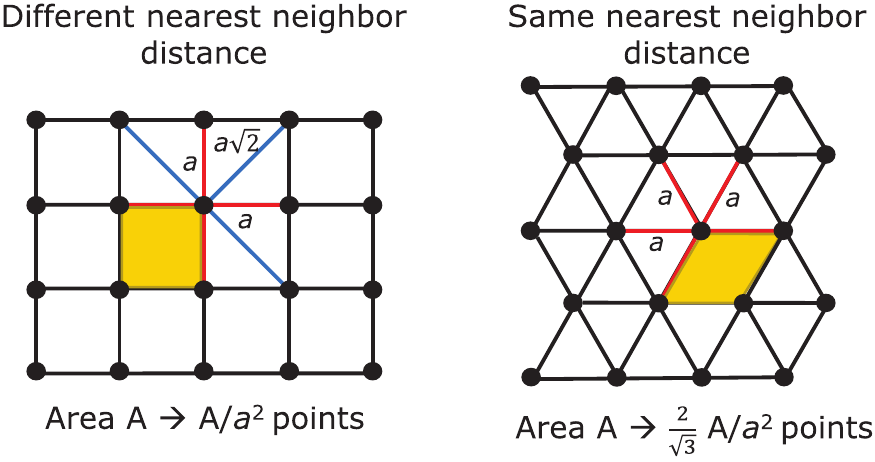}
	\caption{Comparison between sampling on a square (left) and hexagonal (right) lattice. The lattice constant is $a$ and the shaded area represents the unit cell.}
	\label{hex_scan}
\end{figure}

On a square grid with a lattice constant $a$, as shown on the left image in Fig.~\ref{hex_scan}, the diagonal neighbors have a separation of $a\sqrt{2}$. Consequently, the transmitted fields from two spots separated by $a$ or by $a\sqrt{2}$ yield different amounts of overlap. However, in our hexagonal sampling grid as illustrated on the right of Fig.~\ref{hex_scan}, the six nearest neighbors are all at a distance $a$, while the six second neighbors are at a distance $a\sqrt{3}$. 
Critical sampling (within the band limit imposed by the NA) is achieved for the square lattice at 
\begin{equation}
    a_{sq}^2= \frac{\lambda^2}{\pi}(\mbox{NA})^{-2},
\end{equation}
and for the hexagonal lattice at 
\begin{equation}
    a_{hex}^2= \frac{2}{\sqrt{3}}\frac{\lambda^2}{\pi} (\mbox{NA})^{-2} \approx 1.15 \frac{\lambda^2}{\pi} (\mbox{NA})^{-2}.
\end{equation}

Remarkably, near critical sampling the nearest neighbor Airy spots on the hexagonal grid are almost exactly at the Rayleigh criterion 
\begin{equation}
    a_{hex}=0.994 \, a_{Ray} \,\,\,\mbox{ with }\,\,  a_{Ray}=\frac{z_1 \lambda}{2 \pi} (\mbox{NA})^{-1},
\end{equation}
where $a_{Ray}$ is the Rayleigh distance at which the maximum of one spot coincides with the first zero of its neighbor and the field overlap integral vanishes. Here $z_1 \approx 3.83$ is the first zero of the $J_1$ Bessel function. As a consequence, the nearest neighbor spots at critical sampling have almost zero overlap, while the overlap with the second neighbors is also low.
Thanks to this feature of the hexagonal lattice, we reduce spurious correlations between measured transmitted fields and thus measure a more accurate TM when compared to sampling on a square grid with the same spacing $a$. The hexagonal grid also has around $15\%$ higher sampling density because its unit cell is smaller. The choice of a hexagonal lattice scan therefore allows the measurement of a larger fraction of the complete TM at the same level of correlations and consequently represents the statistics of the scattering properties of the sample more faithfully.

Next, we compare the effect of sampling density for both lattices, for simplicity assuming $\mbox{NA}=1$ and a single polarization component. A good measure to gauge these effects is the normalized rank of the basis of sampling modes, which is defined as the number of linearly independent modes divided by the total number of modes in the basis. 
We numerically compute the normalized rank of a basis of 933 spots within a circular area for the hexagonal grid and 931 spots within the same area for the square grid, as a function of the area density of sampling points.
Computationally we find the normalized rank by constructing the matrix of inner products between the modes of the basis and calculating the fraction of its singular values $s$ that are larger than $c \cdot s_{\textrm{rms}}$, where $s_{\textrm{rms}}$ is the RMS (root mean square) singular value and $c=0.05$ is the numerical tolerance factor below which we consider a singular value to be effectively zero. Our results are quite insensitive to the choice of $c$. 
 
\begin{figure}[tb]
	\centering\includegraphics[width=0.67\textwidth]{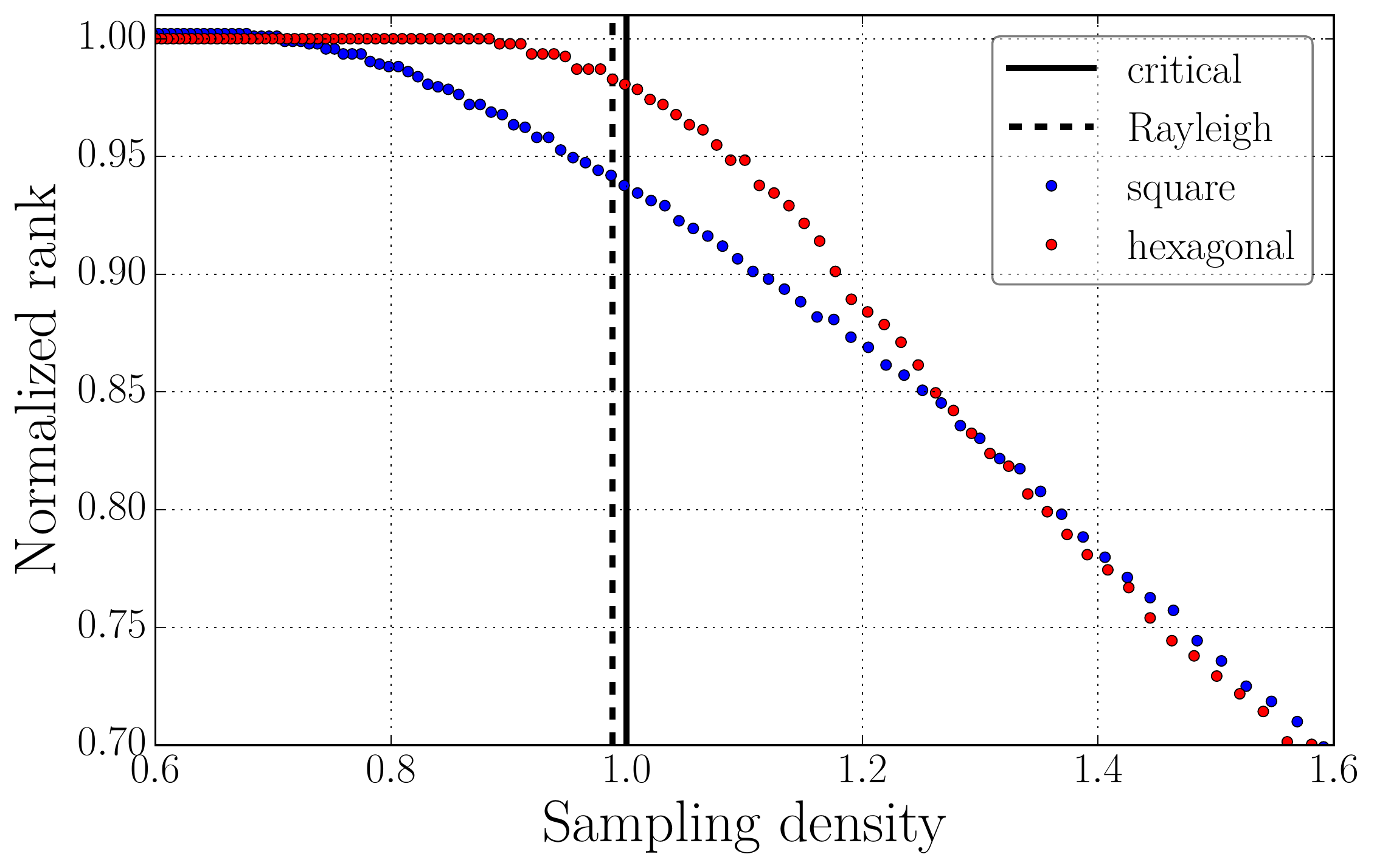}
	\caption{Normalized rank of a sampling basis versus the sampling density for a square and hexagonal lattice. The horizontal axis is normalized to the critical sampling density of $\pi / \lambda^2$, represented by the vertical line at 1. The dashed vertical line indicates the Rayleigh sampling density for the hexagonal lattice. In this numerical calculation $\mbox{NA}=1$. }
	\label{hexsquare}
\end{figure}

Fig.~\ref{hexsquare} shows the normalized rank as a function of the sampling density $N/A$, where $N$ is the number of sampling points, normalized to the critical density of $\pi / \lambda^2$. The vertical line indicates the critical sampling density and the dashed line indicates the Rayleigh criterion for a hexagonal lattice.
Ideally, the spots should be orthogonal and therefore lead to a normalized rank of 1.
This is indeed the case for either lattice if we sample far below the critical sampling density. Close to the critical sampling density, the hexagonal lattice exhibits less spot overlap as indicated by a normalized rank closer to 1. At the critical sampling density, a hexagonal grid yields a normalized rank of 0.981 while a square one gives 0.935. In other words, close to critical sampling the square lattice induces three times more spurious near-zero singular values than the hexagonal one.
Oversampling beyond the critical density can increase signal to noise but only with a square root law in measurement time. Without further data analysis and regularization procedures, it leads to additional spurious zero singular values in the TM as the neighboring lattice spots increase in overlap. Beyond a sampling density of 1.3, both lattices exhibit a similar amount of spurious zeros. This is not necessarily a problem since any critically or oversampled basis may be resampled into a strictly orthogonal set of functions that is complete within the FOV and NA, such as Bessel function modes. Such a resampling procedure requires accurate knowledge of the incident fields and introduces extra parameters and complexity in the data analysis which may be undesirable in real-time applications.
The hexagonal grid of Airy spots spaced at the Rayleigh criterion is almost a complete basis, close to orthogonal and straightforward to implement in an experiment. To maintain symmetry in the system configuration, we choose to use the same basis in the detection plane, yielding a square TM.
In case $\mbox{NA}<1$, the maximum achievable sampling density unavoidably scales with $\mbox{NA}^{2}$ irrespective of the sampling method or lattice, and the results in Fig.~\ref{hexsquare} are still valid with the $x$-axis now representing the scaled sampling density.

The polarization-complete TM can be composed from four measurements of sub-matrices as~\cite{Tripathi12}
\begin{equation}\label{TMpol}
T = 	\begin{pmatrix}
	T_{\mathrm{HH}} & T_{\mathrm{VH}} \\
	T_{\mathrm{HV}} & T_{\mathrm{VV}}
	\end{pmatrix},
\end{equation}

\noindent where H and V are labels for modes which have horizontal and vertical polarization respectively in the back pupil of the objective. Here, the first index indicates the incident polarization while the second indicates the transmitted one. For high NA the polarization is more complicated in the focal plane, but it is uniquely defined in the pupil.

\section{Measurement and data processing}

 \subsection{Apparatus}\label{apparatus}

\begin{figure}[tb]
	\centering\includegraphics[width=0.8\textwidth]{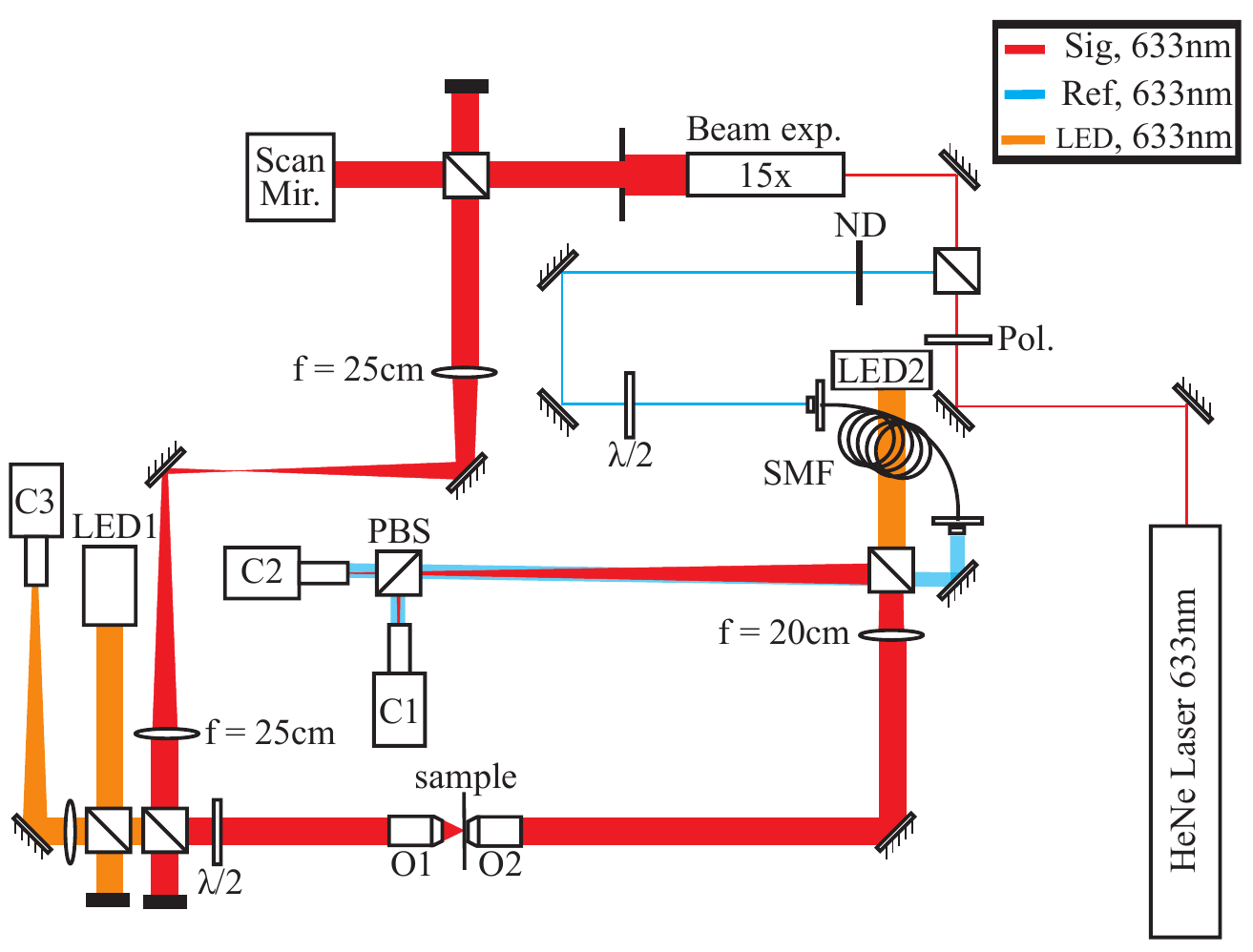}
	\caption{Setup. Light from a Helium-Neon laser is split into a signal and reference beam. The signal (in red) is spatially scanned across a sample with the scan mirror (scan mir.), and for every position on the sample the resulting transmitted scattered light is imaged in real space with two cameras C1 and C2, which measure different output polarizations. The reference beam (in blue) interferes with the signal to retrieve the phase information using off-axis holography. 
	Abbreviations: Pol., polarizer; ND, neutral density filter; beam exp., 15x beam expander; $\lambda/2$, half-wave plate; LED1 and LED2, \SI{633}{\nano\meter} light emitting diodes; O1, 0.95-NA air microscope objective; O2, 1.42-NA oil immersion microscope objective; C3, reflection camera; PBS, polarizing beamsplitter; SMF, single mode fiber.}
	\label{setup}
\end{figure}

Our experimental setup to perform polarization-complete TM measurements is depicted in Fig.~\ref{setup}. Light from a Helium-Neon continuous wave (CW) laser (JDS Uniphase, \SI{5}{\milli\watt}) with a wavelength of \SI{633}{\nano\meter} first passes through a polarizer to ensure a linear and well defined polarization after which it is split into a signal (transmitted) and reference (reflected) beam. The signal beam is expanded, and directed with a two-axis scan-mirror (Newport FSM-300) towards the sample. A 4-$f$ relay telescope images the scan mirror to the pupil of the microscope objective O1 (Zeiss Achroplan 63x, 0.95-NA). A $\lambda/2$-plate on a motorized rotation stage controls the incident beam polarization. The transmitted light is collected with an oil immersion microscope objective O2 (Olympus PlanApo N 60x, 1.42-NA). The vertical and horizontal polarization components of the light are split by a polarizing beamsplitter (PBS) and imaged through a tube lens on two separate cameras C1 and C2 (Basler daA1280-54um) respectively, allowing simultaneous measurements in both polarization channels. The complex fields are retrieved using digital off-axis holography~\cite{Leith:62, Takeda:82, Cuche} by interfering the signal beam with a slightly tilted reference beam. The reference beam is delivered through a single mode polarization-maintaining fiber which ensures a near-Gaussian spatial profile. The reference beam is expanded to overfill the camera chip and provide a nearly flat intensity profile. The signal and reference path lengths are equal to within a few cm, which is crucial to avoid drifts due to changes of the laser frequency and coherence length. The $\lambda/2$-plate in the reference arm controls the ratio of the reference beam power directed to the cameras C1 and C2. Two light-emitting diodes (LED1 and LED2, \SI{633}{\nano\meter}), which are switched off during the measurement, enable wide field illumination to locate desired regions of the sample and to focus the microscope objectives on the sample. 

To measure the polarization-complete TM (eq.~(\ref{TMpol})), we first set the incident polarization to H, after which the scan mirror steers the beam across the front surface of the sample. We start the spot scan at the center and then spiral outwards, which enables straightforward cropping of the resulting matrix to a smaller spatial area. For each incident spot, the cameras C1 and C2 record the V and H polarized transmitted light respectively. The process is repeated for V incident polarization. Dark and reference beam images, averaged over four images to even out the background noise, are taken before and after the measurement.

\subsection{Experimental sampling density}
The effective area covered by the TM measurement is $A=\SI{156}{\micro\meter\squared}$, which corresponds to the hexagonal envelope containing all measured points including half a unit cell outside the outermost spot centers so that the Airy spots belonging to this layer are contained entirely inside the measurement area. This area corresponds to $2\pi A / \lambda^2 = 2443$ optical modes. However, our experimental TM has a dimension of $1838 \times 1838$, resulting in a sampling density of $1838/2443 \approx 0.75$. 

The specified NA of our microscope objective O1 is $0.95$, but the effective NA retrieved by examining the first zero of the Airy spot is somewhat lower and estimated to be $0.9$ possibly due to apertures and slight aberrations in the optical system.
The sampling density is therefore estimated to be at $0.75/0.9^2 = 0.93$ of the maximum sampling density allowed by the NA and close to the Rayleigh criterion at $0.988$.
Hence, according to Fig.~\ref{hexsquare}, we expect to be in a situation of slight undersampling but preserving the statistics of the TM very well with only about 1\% of spurious zero singular values.

\subsection{Digital filtering of the transmitted fields}\label{subsec:Fourier_filter}
The transmitted fields are measured through the optical system with an NA of 1.4 (objective O2), which exceeds the NA of the incident fields. For the case of a clear medium, and in general whenever we want to obtain a square TM, we need the transmitted fields to be filtered to the same NA as the incident fields. We accomplish this by digitally Fourier filtering the measured fields. 
We filter with a supergaussian disc $SG$ with radius $R$, 
\begin{equation}
    SG(r) = \exp\left[- 2 \left(\frac{r}{R} \right)^n \right],
\end{equation}

\noindent where $r$ is the radial coordinate and the supergaussian exponent is chosen to be $n=50$. The supergaussian filter reduces the Gibbs phenomenon~\cite{eldar_2015}.
We set  $R=\SI{1.59}{\micro\meter^{-1}}$, which corresponds to a filter that drops off sharply beyond $\mbox{NA}=0.95$, so that it just passes the spatial frequencies present in the incident beam.

\subsection{Sampling the transmitted field}\label{subsec:airy_sampling}

We convert the measured and filtered fields to a vector by sampling at the positions that correspond to the spot centers of our incident field, which we obtain from the centers of mass (COM) of the transmitted images of the input Airy spots focused on a plain glass slide. The lattice constant of the hexagonal grid of the output basis is the mean separation of the COM of the input spots. We sample the fields at the output lattice points by linearly interpolating the field values from neighboring pixels. In this manner, the output sampling vector has the same size as the incident sampling vector which is equal to the number $N$ of input spots. The measured TM is therefore a square $N \times N$ matrix. 

\begin{figure}[tb]
	\centering\includegraphics[width=0.8\textwidth]{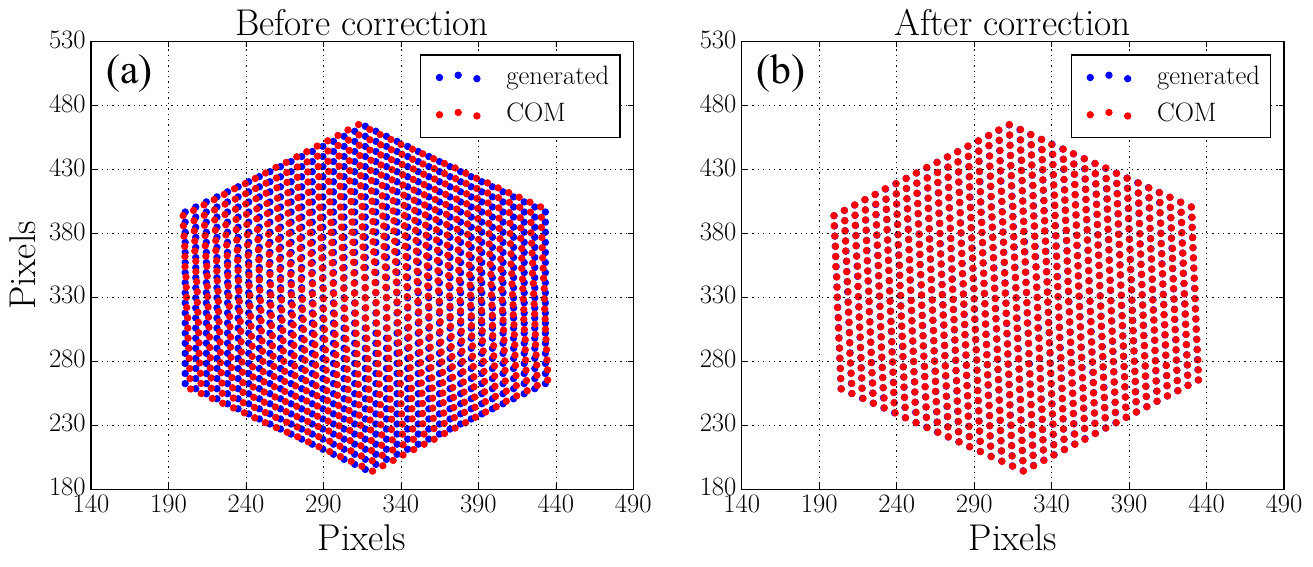}
	\caption{Coordinates of the generated hexagonal grid (blue dots) representing the output sampling basis (a) before and (b) after performing an affine transformation to match the grid found from the centers of mass (COM) of the focused input laser spots (red dots).}
	\label{grid_affine}
\end{figure}

Due to a rotation of the CCDs and possible conformal distortion induced by optical elements, the input and output bases do not overlap exactly. This problem is usually ignored when measuring TMs of strongly scattering samples as there are no theoretical values of the matrix elements to compare to, but it leads to incorrect results for clear samples. To correct for this mismatch with a minimum number of parameters, we use an affine transformation to 
map the input grid to the measured grid of COM points.

A least squares algorithm is used to minimize the difference between the COM coordinates and the transformed grid coordinates. The result of this transformation is shown in Fig.~\ref{grid_affine}. In (a), the generated (output) and measured (input) grids do not initially overlap. After performing an affine transformation on the generated lattice, Fig.~\ref{grid_affine}(b) demonstrates that the input and output grids are superimposed to within \SI{60}{\nano\meter}.

\begin{figure}[tb]
\centering\includegraphics[width=0.8\textwidth]{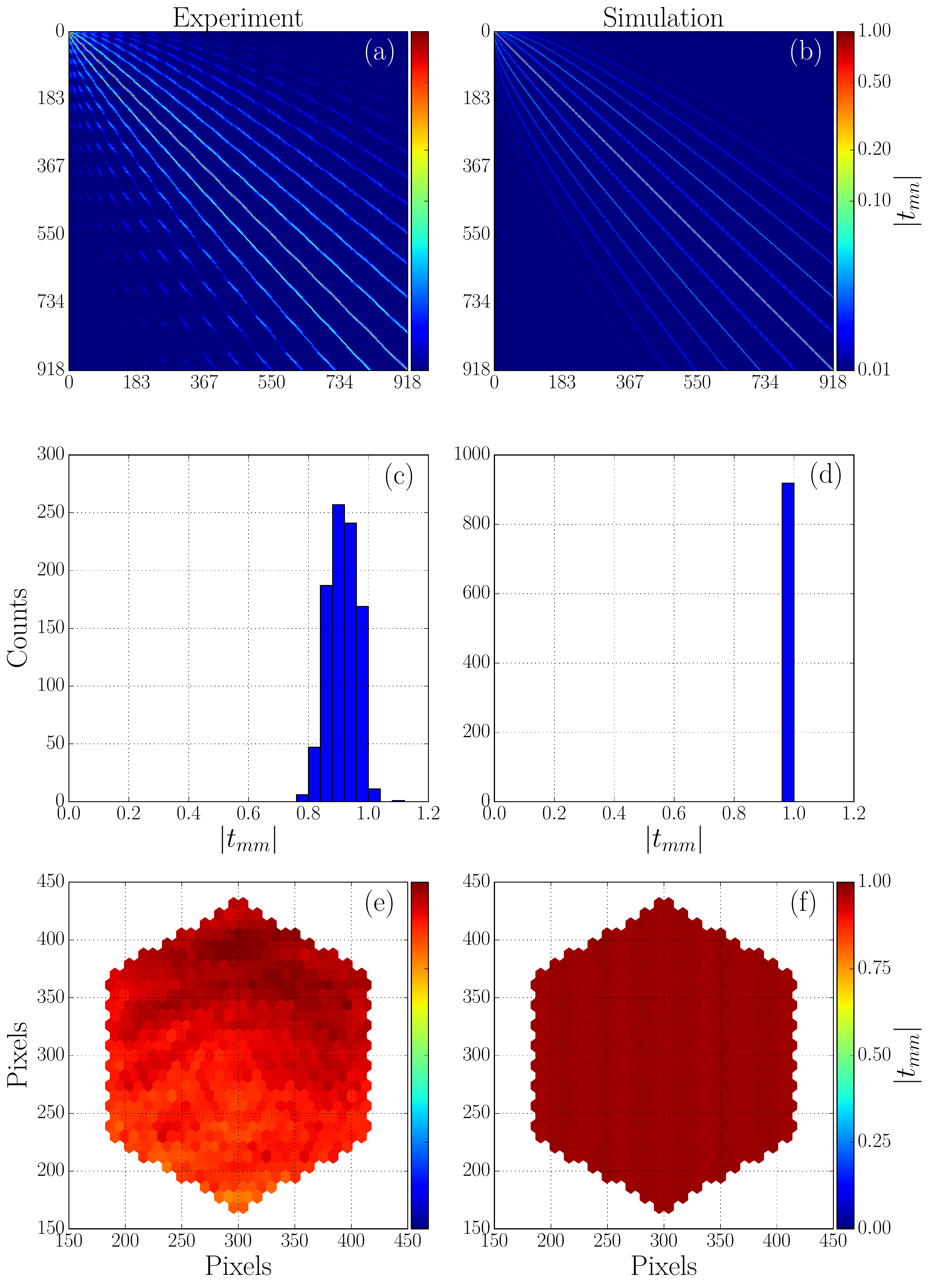}
\caption{Results of the TM measurement and simulations. The left and right column display the measured and simulated results respectively. (a,b) Magnitude of the TM elements for the $919 \times 919$ $T_{\mathrm{HH}}$ polarization component of a zero-thickness reference. (c,d) Histogram of the magnitudes of the diagonal elements. (e,f) Magnitude of the diagonal elements $\lvert t_{mm} \rvert$ as a function of the input position $m$.}
\label{TM}
\end{figure}
\subsection{Phase drift correction}\label{subsec:phase_drift}

During the course of the measurement slow phase drifts occur because of temperature, mechanical and laser wavelength drifts. These phase drifts affect the statistics and reproducibility of the measured TM. A typical measurement of a complete TM of dimension $1838 \times 1838$ elements takes approximately $11$ minutes, and phase drifts of order \SI{1}{\radian} occur in this time. Similar to the method used in~\cite{Ploschner14}, we account for these phase drifts by taking a reference image after every 20 scan positions of the input laser beam. The reference position is chosen to be the central spot of the hexagonal spiral. The phase of these reference fields is then compared to the phase of the starting spot by calculating the angle of their inner product. The phase drifts are interpolated with a one-dimensional smoothing spline fit. Finally, the computed corrections are applied to the retrieved fields to compensate for the phase drifts, reducing the inter-frame phase error to \SI{0.07}{\radian}.

\section{TM of a zero-thickness reference}\label{sec:zero_thickness}

We study the accuracy and precision of our method by measuring the TM of a zero-thickness sample, i.e., objectives O1 and O2 are focused in the same plane. The focal plane we choose is the interface between a crown glass microscope cover slide and air. We note that this zero-thickness TM does not contain any near field terms according to the definition of the scattering matrix, and can consequently be measured with far-field optics.
We scan 919 incident spots, corresponding to 18 complete hexagon layers across the interface, which leads to a polarization-complete TM of $1838 \times 1838$ elements.

A simulated TM is obtained by scanning a numerically generated Airy field on a hexagonal grid, spaced by the Rayleigh criterion just as in the experiment. The number of spots is 919, corresponding to a single polarization component as cross-polarization terms are absent in the numerical case.

The magnitude of one component (H incident, H transmitted) of the measured and simulated TM elements is shown in Fig.~\ref{TM}. The magnitude is normalized to the RMS singular value of $T_{\mathrm{HH}}$. In Fig.~\ref{TM}(a), the main diagonal of the TM is close to 1 and dominates the off-diagonal elements, with 84\% of the total power in the diagonal elements. This indicates that the fields associated to different input spots are nearly orthogonal. Other lines, which are clearly visible due to the logarithmic color scale, emerge from the crosstalk between neighbor and next-neighbor Airy disks. This crosstalk is also visible, albeit at a slightly lower amplitude, in the simulated TM elements in Fig.~\ref{TM}(b). In this case 95\% of the total power is in the diagonal elements, indicating a smaller level of crosstalk with non-nearest neighbor points.
In the histogram in Fig.~\ref{TM}(c) we plot the distribution of the magnitude of the diagonal elements, which is peaked close to 1, as expected for a situation with only small crosstalk. The broadened width of the peak compared to the calculated one in Fig.~\ref{TM}(d) is attributed to experimental noise and imperfect imaging of the scan mirror on the pupil of the objective O1.
In Fig.~\ref{TM}(e) we show the magnitude of the diagonal elements
as a function of their input position. The nonuniform distribution favoring the upper positions of the grid indicates a slightly angle-dependent transmission of our optical system.
In the numerical case displayed in Fig.~\ref{TM}(f), the diagonal elements exhibit very small fluctuations of around $1\%$ which arise due 
to the conversion from a rectangular pixel-based sampling to a hexagonal sampling. We conclude that our experimental data agrees very well with numerical simulations even though this direct comparison detects imperfections in the experiment very sensitively.
 
\begin{figure}[tb]
	\centering\includegraphics[width=0.8\textwidth]{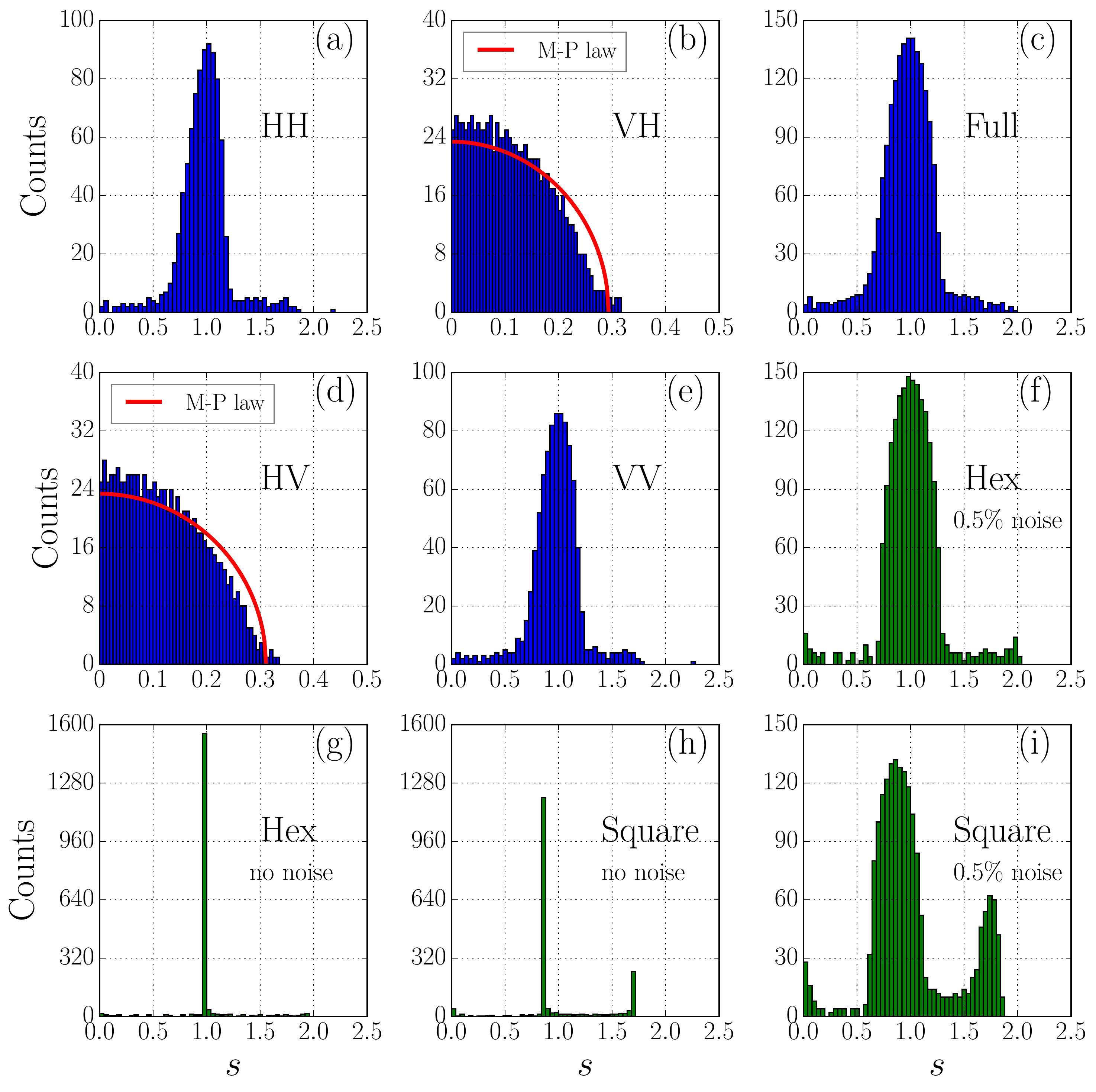}
	\caption{Singular value histograms of the transmission matrix. (a,b,d,e)~Singular values $s$ of the measured polarization sub-matrices and (c)~of the full experimental TM. 
	(f)~Simulated histogram for the full TM including 0.5\% RMS noise and (g)~without noise. (h)~Simulated TM for square lattice sampling with the same lattice constant, without noise and (i)~with noise.	In the cross-polarization matrices (b,d), the red curve corresponds to the Marchenko-Pastur law of a random matrix of the same size.}
	\label{fig:TM_zero_SV}
\end{figure}

A zero-thickness glass sample is the ideal test case to verify whether the sampling procedure leads to distortions of the TM statistics.
An important quantitative probe of these statistics is the distribution of the singular values \cite{pendry1990, Miller19}.
The singular value histograms of the individual polarization sub-matrices and of the complete TM are plotted in Fig.~\ref{fig:TM_zero_SV}. All histograms are normalized to the RMS singular value of the respective co-polarized sub-matrix. In Fig.~\ref{fig:TM_zero_SV}(a,e), associated to polarization conserving measurements, the singular value histograms exhibit a pronounced peak at $s=1$ corresponding to completely transmitting channels. The histogram exhibits a low pedestal extending from $s=0$ to almost $s=2$.
In Fig.~\ref{fig:TM_zero_SV}(b,d), the singular value distributions for the cross-polarization matrices follow the Marchenko-Pastur law~\cite{MP_law}, which predicts the singular value probability function for a random uncorrelated matrix~\cite{mehta2004random, Shen2001}. Since the glass sample does not cause appreciable cross-polarization scattering, we attribute the random values to a combination of camera noise and imperfections in the polarizing beamsplitter.
In Fig.~\ref{fig:TM_zero_SV}(c) we show the singular value histogram for the full experimental TM, and it resembles the $T_{\mathrm{HH}}$ and $T_{\mathrm{VV}}$ sub-matrices. The width is a little broader because of the noise from the cross-polarization channels.
The singular value histogram of a simulated TM including 0.5\% RMS Gaussian random noise on all matrix elements is displayed in Fig.~\ref{fig:TM_zero_SV}(f). We observe that the width of the peak and the pedestal of the experimental histogram are well reproduced by this model. Without noise, as shown in Fig.~\ref{fig:TM_zero_SV}(g), the simulated histogram is much narrower, but the pedestal remains. We conclude that the peak width is caused by experimental noise while the pedestal is a sampling feature. 
The low density of effectively zero singular values ($s<0.05$) indeed matches the results shown in Fig.~\ref{hexsquare}.
For comparison we plot in Fig.~\ref{fig:TM_zero_SV}(h,i) the corresponding histograms for the case of square-lattice sampling with the same lattice constant without and with noise, respectively. The square-lattice sampling introduces much stronger spurious side lobes in the histogram, which obscure the true statistics of the sample TM. This proves clearly the advantage of sampling the TM on a hexagonal grid at the Rayleigh criterion.

\section{Conclusion}

In summary, we have measured the optical transmission matrix (TM) of a zero-thickness medium using a novel sampling method, namely a hexagonal sampling grid of Airy disks at the Rayleigh criterion. We show in numerical simulations that this sampling method induces far less distortions in the singular value statistics than the usual procedure of sampling on a square grid. We provide experimental verification of the statistics in a reference case of a zero-thickness sample. Experimentally verifying that sampling does not lead to distortion of the statistics is an important prerequisite for probing more complex samples such as scattering media.
The advantage of sampling on a Rayleigh-criterion spaced hexagonal grid applies to any kind of waves including ultrasound and microwaves, and it can be used to measure the transmission matrix of fibers~\cite{cizmar,Rosen2015FocusingAS,Descloux2016,Mahalati2013,Borhani2018}, waveguides~\cite{shi2012} and for full scattering matrix measurements of samples in other geometries.

\section*{Funding}
Netherlands Organization for Scientific Research NWO (Vici 68047618)

\section*{Acknowledgments}
The authors thank Sanli Faez and Dries van Oosten for fruitful discussions and Paul Jurrius, Dante Killian and Cees de Kok for technical support.

\section*{Disclosures}

The authors declare no conflicts of interest.


\bibliography{references}


\end{document}